\documentclass[runningheads]{llncs}
\usepackage[T1]{fontenc}
\usepackage{graphicx}
\usepackage{adjustbox}
\usepackage{caption}
\usepackage{subcaption}
\usepackage{amsmath}
\usepackage{amssymb}
\usepackage{times}
\usepackage{epsfig}
\usepackage{threeparttable}
\usepackage{multirow}
\usepackage{hyperref}
\usepackage{color}
\begin{document}
%
\title{Feature Representation Learning for Robust Retinal Disease Detection from Optical Coherence Tomography Images}
\titlerunning{RFRL-Network for Robust Retinal Disease Detection}
%
\author{
Sharif Amit Kamran\inst{1} \and
Khondker Fariha Hossain\inst{1} \and
Alireza Tavakkoli\inst{1} \and 
Stewart Lee Zuckerbrod\inst{2} \and 
Salah A. Baker\inst{3}}

%
\authorrunning{Kamran et al.}
%
\institute{Dept. of Computer Science \& Engineering, University of Nevada, Reno, NV, USA\and Houston Eye Associates, TX, USA\and School of Medicine, University of Nevada, Reno, NV, USA\\ }
%

%
\maketitle              
\begin{abstract}
Ophthalmic images may contain identical-looking pathologies that can cause failure in automated techniques to distinguish different retinal degenerative diseases. Additionally, reliance on large annotated datasets and lack of knowledge distillation can restrict ML-based clinical support systems' deployment in real-world environments. To improve the robustness and transferability of knowledge, an enhanced feature-learning module is required to extract meaningful spatial representations from the retinal subspace. Such a module, if used effectively, can detect unique disease traits and differentiate the severity of such retinal degenerative pathologies. In this work, we propose a robust disease detection architecture with three learning heads, i) A supervised encoder for retinal disease classification, ii) An unsupervised decoder for the reconstruction of disease-specific spatial information, and iii) A novel representation learning module for learning the similarity between encoder-decoder feature and enhancing the accuracy of the model. Our experimental results on two publicly available OCT datasets illustrate that the proposed model outperforms existing state-of-the-art models in terms of accuracy, interpretability, and robustness for out-of-distribution retinal disease detection.

\end{abstract}
\section{Introduction}
\label{sec:introduction}
Diabetes affects up to 10.2\% of the population globally, and it is projected to grow to 783 million people by 2045 \cite{sun2022idf}. With its prevalence, one-third of every diabetic patient develops Diabetic Retinopathy \cite{ting2016diabetic}.  As the disease progresses, it can lead to Diabetic Macular Edema (DME), which is caused by damaged blood vessels which leak fluid and cause swelling, resulting in blurry vision. DME is a leading cause of vision loss among the working-age population of most developed countries \cite{lee2015epidemiology} and affects approximately 750,000 people in the US.  Even though, significant advancements in the anti-VEGF (vascular endothelial growth factor) therapy has provided patients with treatment options that can delay the progression of the degeneration \cite{lim2012age}, without early detection of this disease can result in permanent vision loss. 

Optical Coherence Tomography (OCT) is an imaging procedure used where back-scattered light is projected for capturing and analyzing the sub-retinal layers and any deformities, aneurysms, or fluid build-up \cite{sri2014}. OCT images are manually examined by expert ophthalmologists to diagnose any underlying retinal degenerative diseases. Hence, miscategorization of diseases can happen due to human error while performing the differential diagnosis. The underlying reason is the stark similarity between DME and other retinal degenerative neuro-ocular diseases such as Age-related Macular Degeneration (AMD), choroidal neovascularization (CNV), or Drusen  \cite{yau2012global}. For instance, in wet-AMD leaky blood vessels grow under the retina and cause blurry vision similar to DME. On the other hand, in choroidal neovascularization (CNV), which is a late stage of AMD, new blood vessels grow from the Bruch membrane (BM) into the subretinal pigment epithelium (sub-RPE). Experts usually encounter problems while differentiating between DME and AMD. 

Recently, with the advent of deep learning, many automated systems have been deployed for the early detection of retinal degenerative diseases. Also, these architectures are trained and tested on the same data distribution and resulting in high prediction accuracy in their respective tasks. However, if applied to the out-of-distribution datasets, the model fails to capture intrinsic features to accurately classify the underlying degenerative condition. So, the robustness and knowledge distillation of such systems are contentious. To address this problem, we propose a novel supervised-unsupervised representation learning module that can improve the accuracy of any retinal disease classification model on unseen data distribution. Moreover, this module can be attached to any pre-trained supervised image classification models. Our extensive qualitative and quantitative experiments illustrate the proposed module's interpretability, robustness, and knowledge transferability. 

\section{Related Work}
Many image processing techniques have been proposed to diagnose retinal degenerative diseases One proposed method is to segment, fuse and delineate multiple retinal boundaries to detect anomalies and diseases from Retinal OCT images \cite{debuc2011review}. Graph cuts and region-based delineation methods have also been proposed to detect different abnormalities and degeneration in the retinal subspace ~\cite{vermeer2011automated} and for diagnosing the thickness of the choroidal folds and neovascularization~\cite{alonso2013automatic,philip2016choroidal}. Early approaches identified Diabetic Macular Edema (DME) with 75-80\% sensitivity score~\cite{sanchez2004retinal,ege2000screening}. In contrast, segmentation-based approaches can help diagnose underlying causes of liquid buildup in the subretinal layers by detecting irregular retinal features and comparing the differences between healthy and the degenerated retinal tissue~\cite{meindertniemeijer2012,quellec2010three,lee2010segmentation}. Even though segmentation approaches have shown success, it results in severe inaccuracies when applied to OCT images acquired from different OCT acquisition systems ~\cite{kafieh2013review}.

Deep convolutional neural networks (DCNNs) have recently received state-of-the-art results in identifying different retinal degenerative diseases~\cite{lee2017deep}. For example, Fang et al. incorporated CNNs with graph search to simultaneously segment retinal layer boundaries and detect degeneration for patients having Age-related Macular Edema\cite{fang2017automatic}. On the other hand, Xu et al. proposed a Dual-stage framework that utilizes CNNs to segment retinal pigment epithelium detachment \cite{xu2017dual}. Kamran et al. proposed a novel deep learning architecture called OpticNet-71 that achieved state-of-the-art accuracy in two OCT image benchmarks for identifying DME, AMD, CNV, etc\cite{kamran2019optic}. Subsequent works have also utilized MobileNet-v2 \cite{nugroho2021convolutional}, MobileNet-v3 \cite{wang2022classification} and VGG16 \cite{kim2021retinal,subramanian2022classification} for retinal disease classification.  Despite achieving high accuracy in their respective benchmarks, most of these architectures do not converge and lack robustness and knowledge transferability when evaluated on an out-of-distribution dataset, as reported in \cite{kamran2020improving}. To alleviate this, the authors in \cite{kamran2020improving} proposed a joint-attention network with a supervised classifier and unsupervised image reconstruction module. Adopting an adaptive loss function, the model achieved 1.8-9.0\% improvement over three baseline state-of-the-art models, namely ResNet-50, MobileNet-v2, and OpticNet-71 on an out-of-distribution dataset. However, the model can perform poorly due to adaptive learning prioritizing image reconstruction over disease classification. Moreover, due to using non-learnable upsampling layers in the image reconstruction module, the model does not retain intrinsic features, hampering overall accuracy and robustness. 

We propose a novel robust feature representation learning (RFRL) network that can be incorporated into any deep classification architecture for robust out-of-distribution retinal disease detection. Our module consists of 1) a supervised learning head for classifying diseases, 2) an unsupervised decoder head for disease-specific spatial image reconstruction, and 3) a novel representation learning head for finding similar features robustly from the encoder and decoder of the architecture. Furthermore, the proposed representation learning head incorporates a novel multi-stage feature similarity loss to boost the model's accuracy on out-of-distribution samples. Our experiments confirm that the proposed RFRL network incorporated into baseline and state-of-the-art architectures significantly improves accuracy, sensitivity, and specificity for OOD datasets. Furthermore, we qualitatively evaluate its interpretability using GradCAM\cite{selvaraju2017grad} and GradCAMv2\cite{chattopadhay2018grad} to prove its clinical significance. Expert ophthalmologists can leverage this module to improve disease detection from OCT b-scans on OOD datasets and avoid sub-par performance.
\begin{figure}[t]
    \centering
    \includegraphics[width=1\columnwidth]{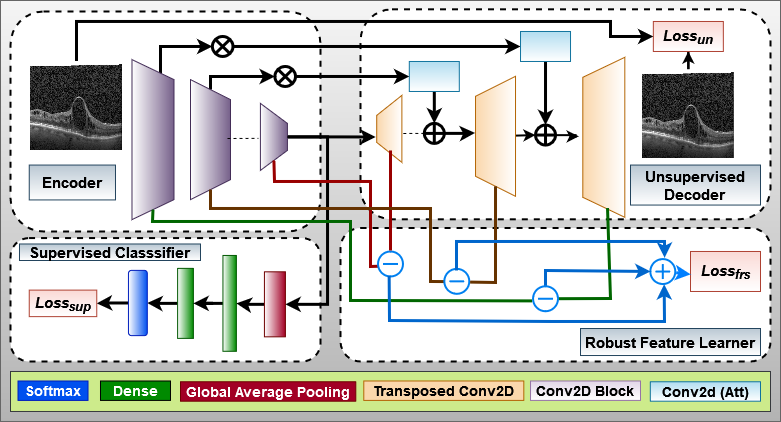}
    \caption{RFRL-network consisting of (1) Encoder, (2) Unsupervised Decoder, (3) Robust Feature Learner and (4) Supervised Classifier. The unsupervised decoder incorporates mean-squared-error (MSE), supervised classifier utilizes cross-entropy loss and the robust feature learner adopts our novel Feature Representation Similarity (FRS) loss.}
    \label{fig1}
\end{figure}

\section{Methodology}

\subsection{Robust Feature Learning Architecture}
An image classification architecture consists of an encoder module with multiple learnable convolution layers (top left in Fig.~\ref{fig1}) and a supervised downstream classification head with global average pooling and dense layers (bottom left in Fig.~\ref{fig1}). In addition, we propose an unsupervised decoder head  (top right in Fig.~\ref{fig1}) for image reconstruction and a feature representation learning head between the encoder and decoder (bottom right in Fig.~\ref{fig1})to be incorporated with the architecture. The objective of the unsupervised module is to reconstruct the original image and learn its intrinsic spatial features. Moreover, we use skip-connections to retain disease-specific and domain-invariant spatial features that help in overall classification performance. 

Joint-Attention Network \cite{kamran2020improving} incorporated non-learnable bilinear upsampling layers in their unsupervised decoder. As a result, the spatial feature information was insufficient to reconstruct the original image with disease pathology. However, their experiment illustrated an improved overall accuracy over the baseline methods. To further improve upon this, we propose learnable transposed convolution layers for upsampling in our unsupervised decoder, as shown in Fig.~\ref{fig1}. The transposed convolution consists of kernel size, $k=3$, and stride, $s=2$.  Consequently, our method can retain class-specific and domain-independent spatial salient information, an essential missing feature in the low performance of the traditional encoder-based architectures in practice. 

The improved decoder utilizes a mean-squared-error loss to learn original image features in an unsupervised manner without any ground truth or class labels. Outputs from each downsampling block in the encoder module are connected with each upsampling block in the decoder module with a skip connection. A convolution layer is used as an attention mechanism and element-wise summation of features between encoder and decoder module, as illustrated in Fig.~\ref{fig1}. The output of the decoder can thus be defined by the Eq.~\eqref{eq1}:
\begin{equation}
   O_{recon} = E_0\otimes A_n \oplus D_{n}(\cdots E_{n-2} \otimes A_2 \oplus D_1 (E_{n-1}\otimes A_1 \oplus D_0 (E_n\otimes A_0)))
   \label{eq1}
\end{equation}
where, $E_0,E_1,...E_n$ are the output of each of the downsampling blocks, while $D_0,D_1,..D_n$ symbolise the transposed convolution layers of the decoder. $A_0,A_1,...,A_n$ are the convolution operations of attention skip-connections for maintaining the same depth as the output of the corresponding up-sampling layer for element-wise summing.

\subsection{Proposed Representation Learning Loss}
In order to extract intrinsic and robust features, we propose a novel feature representation similarity loss, which is visualized in Fig.~\ref{fig1}. The loss calculates the similarity between each of the down-sampling and up-sampling blocks successively, given in Eq.~\ref{eq2}. 
\begin{equation}
   \mathcal{L}_{frs} = \frac{1}{N} \sum_{i=0}^{n}  \Vert E_{i}(x) - D_{n-i}(x)  \Vert 
   \label{eq2}
\end{equation}
where $x$ is the input image, $E(x)$ is the feature output of the downsampling blocks, and $D(x)$ is the output of the upsampling blocks. The summation is divided with $N$, which is the number of upsampling or downsampling layers for calculating the mean.

\subsection{Final Objective Function}
Our architecture incorporates two other loss schemes to make the learning robust and interpretable. First, for the classification of different retinal diseases, we use Categorical cross-entropy in the supervised classifier given by $\mathcal{L}_{sup}$ equation in Eq.~\eqref{eq3}. Secondly, for calculating the difference between real and reconstructed images, we use  Mean-Squared Error (MSE) loss in the unsupervised decoder given by the $\mathcal{L}_{un}$ in Eq.~\eqref{eq4}.  

\begin{equation}
    \mathcal{L}_{sup} = -\sum_{i=1}^{c} y_i\log(y'_i)
    \label{eq3}
\end{equation}

\begin{equation}
    \mathcal{L}_{un} = \frac{1}{M}\sum_{i}^{M}(x_i - x'_i)^2 
    \label{eq4}
\end{equation}

Here, in Eq.~\ref{eq3}, $y_i$ signifies the ground truth label, $y'_i$ symbolizes predicted output, $c$ is for the number of disease categories. In Eq.~\ref{eq4}, $x_i$ is the input image, $x'_i$ is the reconstructed output image, and $M$ is the number of pixels in the image. 

By combining Eq.~\ref{eq2}, Eq.~\ref{eq3} and Eq.~\ref{eq4}, we create our final objective function given in Eq.~\ref{eq5}.

\vspace{-.1in}
\begin{equation}
     \mathcal{L} = \mathcal{L}_{sup} + \mathcal{L}_{un} + \mathcal{L}_{frs}
    \label{eq5}
\end{equation}
Unlike the Joint-attention-Network \cite{kamran2020improving} we do not utilize any weights to prioritize one or more of these losses. Instead, equal priority is given to all of them for robust feature learning and improved accuracy, sensitivity, and specificity.

\section{Experiments}
\subsection{Data-set Processing}
We evaluate our proposed architecture models on two separate data-sets, \textbf{Srinivasan2014}~\cite{sri2014} and \textbf{OCT2017}~\cite{kermany2018identifying}. The \textbf{Srinivasan2014} dataset comprises 3,231 images, out of which 2,916 are for training and 5-fold cross-validation, and 315 are for testing. It has three categories of images, Normal, AMD, and DME. The model with the best validation result on the \textbf{Srinivasan2014} dataset was used for further testing on the out-of-distribution (OOD) second dataset, i.e., \textbf{OCT2017}. The \texttt{OCT2017} consists of four distinct categories of 1000 test images. We take 250 cases of Normal and DME (in total, 500 samples) for OOD testing. All images for training and testing were resized to $224\times224\times3$ resolution. Moreover, we incorporated random data augmentation techniques such as horizontal flip, rotation, zoom, width, and height shift for the training set. 

\subsection{Hyper-parameter Tuning}
For training baseline methods and RFRL-Networks, we used Adam optimizer. Moreover, it was the same for both the supervised classifier and the unsupervised decoder. The initial learning rate, $lr= 0.0001$. We utilized a mini-batch of $b=4$ and trained all methods for 50 epochs. We reduced the learning rate by $0.1$ if the validation loss did not decrease for six epochs. he code repository is provided in this \href{https://github.com/SharifAmit/RFRL-Net}{link}.

\subsection{Performance Metrics}
We used three standard metrics for calculating the Accuracy, Sensitivity (True Positive Rate)  and Specificity (True Negative Rate). The metrics are caclulated as follows, Accuracy $= \frac{1}{N}\sum\frac{TP+TN}{TP + TN + FN + FP}$, Sensitivity $=\frac{1}{K}\sum\frac{TP}{TP + FN}$, and Specificity $=\frac{1}{K}\sum\frac{TN}{TN + FP}$.

\begin{table}[!t]
\caption{Test Results on In-distribution Srinivasan2014\cite{sri2014} Dataset}
\begin{adjustbox}{width=1\linewidth,center}
\begin{tabular}{c|c|cc|cc|cc }\hline
\textbf{Architectures}&
\textbf{Year} & 
\multicolumn{2}{c|}{\textbf{Accuracy}} &
\multicolumn{2}{c|}{\textbf{Specificity}} & \multicolumn{2}{c}{\textbf{Sensitivity}} \\ \hline\hline
ResNet50-v1 \cite{serener2019dry}& 2018 &
\multicolumn{2}{c|}{94.92}    & \multicolumn{2}{c|}{97.46}    & \multicolumn{2}{c}{94.92}    \\ 
OpticNet-71 \cite{kamran2019optic}& 2019 &
\multicolumn{2}{c|}{100.00}      & \multicolumn{2}{c|}{100.00}         & \multicolumn{2}{c}{100.00}         \\ 
MobileNet-v2 \cite{kamran2020improving,nugroho2021convolutional}& 2020 & \multicolumn{2}{c|}{97.46}& \multicolumn{2}{c|}{98.73}& \multicolumn{2}{c}{97.46}\\ 
VGG16 \cite{subramanian2022classification,kim2021retinal}& 2021 & \multicolumn{2}{c|}{99.04}& \multicolumn{2}{c|}{99.52}& \multicolumn{2}{c}{99.04}\\
MobileNet-v3 \cite{wang2022classification}& 2022 & \multicolumn{2}{c|}{83.80}& \multicolumn{2}{c|}{91.90}& \multicolumn{2}{c}{87.61}\\
\hline
Joint-Attention-Network ResNet50-v1 \cite{kamran2020improving} & 2020 & 
100 & $\uparrow$5.08  & 100.00 & $\uparrow$2.54 & 100.00 & $\uparrow$5.08 \\
Joint-Attention-Network OpticNet-71 \cite{kamran2020improving}  & 2020 &
99.68 &$\downarrow$0.32    & 99.84&$\downarrow$0.16& 99.68 &$\downarrow$0.32        \\ 
Joint-Attention-Network MobileNet-v2 \cite{kamran2020improving} & 2020 & 
99.36 &   $\uparrow$1.90   & 99.68   & $\uparrow$0.95       & 99.36 &$\uparrow$1.90            \\ \hline
\textbf{RFRL-Network ResNet50-v1} & 2022 & 100 & $\uparrow$5.08  & 100.00 & $\uparrow$2.54 & 100.00 & $\uparrow$5.08 \\
\textbf{RFRL-Network OpticNet-71} & 2022 & 100.0 & $(-)$ & 100.0 & $(-)$& 100.0 & $(-)$ \\
\textbf{RFRL-Network MobileNet-v2} & 2022 & 99.68  & $\uparrow$ 2.22 & 99.84 & $\uparrow$ 1.11 & 99.52 & $\uparrow$ 2.06  \\ 
\textbf{RFRL-Network VGG16} & 2022 & 99.68 & $\uparrow$ 0.64 & 99.84 & $\uparrow$ 0.32 & 99.52 & $\uparrow$ 0.64 \\
\textbf{RFRL-Network MobileNet-v3} & 2022 & 99.36 & $\uparrow$ 15.56 & 99.68 & $\uparrow$  7.78 & 99.52 & $\uparrow$ 11.91 \\ \hline
\end{tabular}
\end{adjustbox}
\label{table1}
\end{table}

\subsection{Quantitative Evaluation}

We worked with five baseline architectures across two distinct data sets to evaluate our model's initial performance. Each of these methods have already been incorporated for retinal disease classification from OCT B-scans \cite{serener2019dry,kamran2019optic,nugroho2021convolutional,subramanian2022classification,wang2022classification}, out of which OpticNet-71 \cite{kamran2019optic} has achieved state-of-the-art result on OCT2017 \cite{kermany2018identifying} and Srinavasan2014 \cite{sri2014} datasets. Subramaniam et al. proposed an architecture based on VGG16 \cite{subramanian2022classification} which achieved superior results on seven different pathologies, namely, AMD, CNV, DRUSEN, DMR, DR, MH, and CSR. Quite recently, Wang et al. proposed a model based on MobileNet-v3 \cite{wang2022classification} which achieved scores on par with Optic-Net on the OCT2017 dataset. For a fair comparison, we trained these five models from scratch on the \textbf{Srinivasan2014} dataset with 5-fold cross validation. After choosing the best model, we tested on the data distribution familiar to the architecture, which is the \textbf{Srinivasan2014} test set of 315 images. We then train, validate and test in the same manner with our proposed RFRL-network on the same data distribution. The quantitative comparison is given in Table.~\ref{table1}. As it can be seen, our method's performance for the five models exceeds the baseline scores. Moreover, Joint-Attention-Network \cite{kamran2020improving} also supersedes the baseline methods; however, both our methods for OpticNet-71 and MobileNet-v2 achieve better scores. For our next benchmark, we test the models on the out-of-distribution test set to evaluate their robustness and knowledge transferability. We use the 500 test images with Normal and DME categories from the \textbf{OCT2017} data-set for this evaluation. As illustrated in Table.~\ref{table2}, our model retains intrinsic spatial information that helps it achieve higher accuracy than the baseline methods and Joint-Attention-Networks. It should be noted that none of the \textbf{OCT2017} images were used for training or validating the models. The most significant improvement is seen in ResNet50-v1, with a 4.0\% increase in accuracy over Joint-Attention-Network.  Also, there was a slight specificity drop for MobileNet architectures. Still, it is negligible, as correctly classifying diseases (sensitivity) is more important than misclassifying patients without conditions (specificity). We only report test results for ResNet50-v1, OpticNet-71, MobileNet-v2 versions of the Joint-attention-network \cite{kamran2020improving} as these were provided in the literature. Nonetheless, the proposed RFRL-network retains more robust and intrinsic features across different methods and evaluation settings.

\begin{table}[!tp]
\caption{Test Results on out-of-distribution OCT2017\cite{kermany2018identifying} Dataset}
\label{tab:my-table}
\begin{adjustbox}{width=1\columnwidth,center}
\begin{tabular}{c|c|cc|cc|cc }\hline
\textbf{Architectures}& \textbf{Year}&  \multicolumn{2}{c|}{\textbf{Accuracy}} & \multicolumn{2}{c|}{\textbf{Specificity}} & \multicolumn{2}{c}{\textbf{Sensitivity}} \\ \hline\hline
ResNet50-v1 \cite{he2016deep} & 2018 & 83.40&$\downarrow$11.52    & 89.40 &  $\downarrow$8.06  & 83.40&$\downarrow$11.52    \\ 
OpticNet-71 \cite{kamran2019optic}& 2019 & 
74.40 &$\downarrow$25.60 & 85.60 & $\downarrow$14.40 & 74.40 &$\downarrow$25.60         \\ 
MobileNet-v2 \cite{sandler2018mobilenetv2}& 2020 & 93.80&$\downarrow$3.66& 96.70&$\downarrow$2.03 & 93.80& $\downarrow$3.66\\
VGG16 \cite{subramanian2022classification,kim2021retinal}& 2021 & 92.40 & $\downarrow$ 6.64 & 95.13 &$\downarrow$ 4.39 & 92.40 & $\downarrow$ 6.64  \\
MobileNet-v3 \cite{wang2022classification}& 2022 & 71.60 & $\downarrow$ 12.20 
& 85.50 & $\downarrow$ 6.40 & 71.60 &  $\downarrow$ 16.01\\
\hline
Joint-Attention-Network ResNet50-v1 \cite{kamran2020improving}  & 2020 &
92.40 & $\uparrow$9.0  & 95.00 & $\uparrow$5.6 & 92.40 & $\uparrow$9.0 \\ 
Joint-Attention-Network OpticNet-71 \cite{kamran2020improving}  & 2020 &
77.40 &$\uparrow$3.0  & 89.00&$\uparrow$3.4& 77.40 &$\uparrow$3.0        \\ 
Joint-Attention-Network MobileNet-v2 \cite{kamran2020improving} & 2020 &
95.60 &   $\uparrow$1.8 & 97.1  & $\uparrow$0.4       & 95.60 &$\uparrow$1.8            \\ 
\hline
\textbf{RFRL-Network ResNet50-v1} & 2022 & 96.40 & $\uparrow$ 13.0 & 97.67 & $\uparrow$ 8.27 & 96.40 & $\uparrow$ 13.0 \\
\textbf{RFRL-Network OpticNet-71} & 2022 & 77.60  & $\uparrow$ 3.2 & 86.20 & $\uparrow$ 0.6 & 77.60 & $\uparrow$ 3.2 \\
\textbf{RFRL-Network MobileNet-v2} & 2022 & 95.80 & $\uparrow$ 2.0 & 95.80 & $\downarrow$ 0.9 & 95.80 & $\uparrow$ 2.0 \\ 
\textbf{RFRL-Network VGG16} & 2022 & 96.80 & $\uparrow$ 4.4 & 97.93 &$\uparrow$ 4.8 & 96.80 & $\uparrow$ 4.4 \\
\textbf{RFRL-Network MobileNet-v3} & 2022 & 74.60 & $\uparrow$3.0 & 84.06 & $\downarrow$ 1.44 & 74.60 & $\uparrow$3.0\\ \hline
\end{tabular}
\label{table2}
\end{adjustbox}
\end{table}

\begin{figure}[t]
    \centering
    \includegraphics[width=0.9\columnwidth]{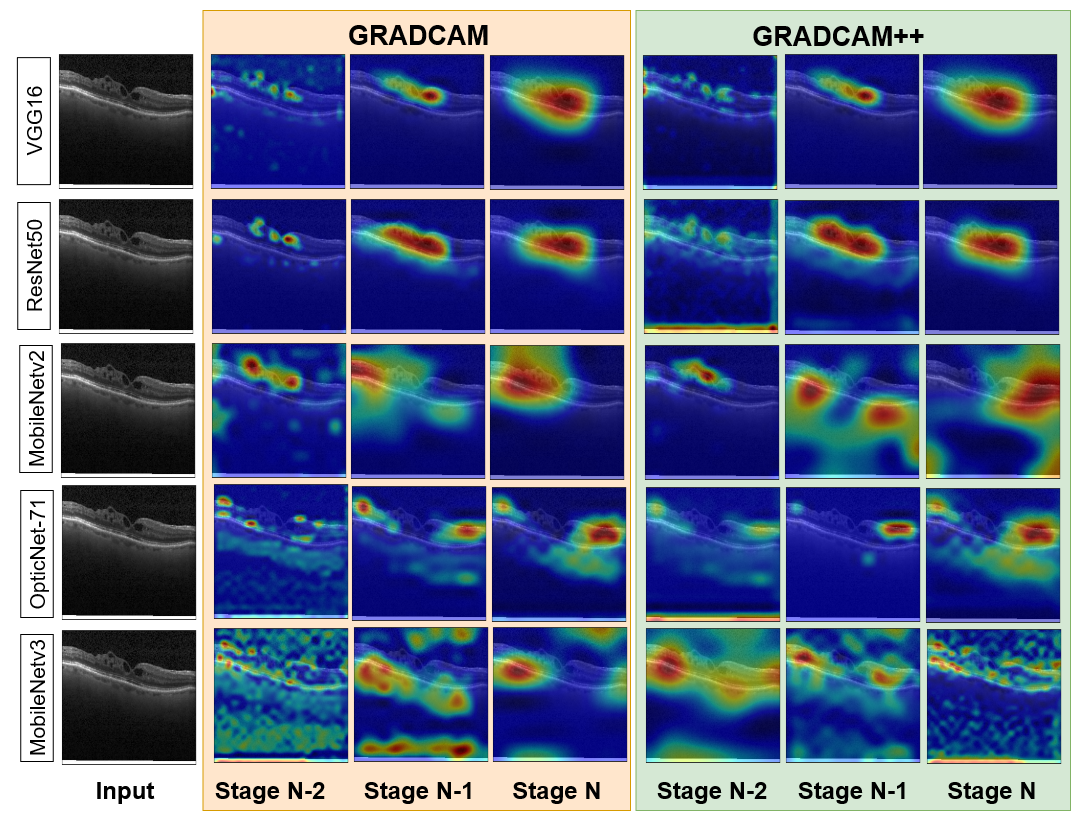}
    \caption{Visualization of back-propagated gradient acitavation signals using GRAD-CAM \cite{selvaraju2017grad} and GRAD-CAM++ \cite{chattopadhay2018grad} on five different RFRL network. Here, Stage N is the last layer of the encoder, the Stage N-1 and Stage N-2 are preceding layers.}
    \label{fig2}
\end{figure}
\subsection{Qualitative Evaluation}
For producing "visual explanations" for judgments made by our CNN architectures for accurate classification, we use GRAD-CAM \cite{selvaraju2017grad} and GRAD-CAM++ \cite{chattopadhay2018grad}. These methods use back-propagated gradients to visualize essential regions of the images of a specific layer, amplified for the classification decision's maximum probability. In Fig.~\ref{fig2}, we illustrate the differences in activations of three stages of the encoder layer for five of our methods on a DME image from OCT2017 \cite{kermany2018identifying} dataset. Stage N is the last convolution layer before the global average pooling, so it does not have any skip-connection with the unsupervised decoder. However, Stage N-1 and Stage N-2 are the previous encoder layers with attention skip-connections. Additionally, they are utilized for Feature Representation Similarity loss, $\mathcal{L}_{frs}$. From Fig.~\ref{fig2}, row 1-2, it is apparent that VGG16 and ResNet50-v1 got activated signals in regions with fluid buildup and hard exudates, explaining the identification of DME in different stages of the encoder. However, in rows 4-5, Optic-Net-71 and MobileNet-v3 got fewer activated signals and activations in the unimportant region, which helped classify the image as DME. The visualization also follows similar trends to standard metrics, where VGG16 achieves the highest to MobileNet-v3, achieving the lowest accuracy in performance. This qualitative visualization helps with our model's overall explainability and knowledge transferability.


\section{Conclusion and Future Work}
\label{sec:conclution}
In this paper, we propose RFRL-network that combines supervised, unsupervised, and feature representation learning to make the robust classifiers for out-of-distribution retinal degeneration detection. Moreover, by incorporating a novel feature representation learning loss, our architecture retains intrinsic and essential feature information that helps with knowledge transferability and explainability. In the future, we wish to extend our work to identifying other retinal degenerative conditions. This can help clinicians in conducting complex differential diagnoses.

\bibliographystyle{splncs04}
%
\bibliography{reference}

\end{document}